# Learning broken symmetries with approximate invariance

Seth Nabat,[1] Aishik Ghosh,[2,3] Edmund Witkowski,[4] Gregor Kasieczka,[5] and Daniel Whiteson[2]

[1]*Taft Charter High School STEAM Magnet, Woodland Hills, California 91364, USA*
[2]*Department of Physics and Astronomy, University of California, Irvine, California 92697, USA*
[3]*Physics Division, Lawrence Berkeley National Laboratory, Berkeley, California 94720, USA*
[4]*Independent Researcher*
[5]*Institut für Experimentalphysik, Universität Hamburg, Hamburg, Germany*

(Received 7 January 2025; accepted 13 March 2025; published 3 April 2025)

Recognizing symmetries in data allows for significant boosts in neural network training, which is especially important where training data are limited. In many cases, however, the exact underlying symmetry is present only in an idealized dataset, and is broken in actual data, due to asymmetries in the detector, or varying response resolution as a function of particle momentum. Standard approaches, such as data augmentation or equivariant networks fail to represent the nature of the full, broken symmetry, effectively overconstraining the response of the neural network. We propose a learning model which balances the generality and asymptotic performance of unconstrained networks with the rapid learning of constrained networks. This is achieved through a dual-subnet structure, where one network is constrained by the symmetry and the other is not, along with a learned symmetry factor. In a simplified toy example that demonstrates violation of Lorentz invariance, our model learns as rapidly as symmetry constrained networks but escapes its performance limitations.

## I. INTRODUCTION

The discovery of new particles or forces in experiments requires the analysis of large volumes of high-dimensional data typically produced by detector systems that surround the interaction point. At the Large Hadron Collider, for example, collisions occur at a rate of 40 MHz and each collision produces millions of sensor outputs. Historically, these data have been reduced using human designed summary features, but recently machine learning models have demonstrated the capacity to extract more information than is preserved in such handcrafted summaries [1–3].

Learning statistical patterns in high-dimensional data space often requires very large volumes of training data to optimize the many model parameters, presenting logistical and economical barriers [4]. A powerful technique for learning with a limited training sample is to leverage symmetries present in the task and exhibited by the data. One approach is to constrain the space of possible solutions to those which respect the symmetry. If, for example, the data are closed under some transformation, the network should learn a function which is likewise invariant (or at least equivariant) under that transformation. This might be done either by incorporating the symmetry directly into the structure of the network, explicitly constraining the functional space that the network can describe and simplifying the learning task, or by reducing the network inputs to a set of invariants, such that the network output is inherently unaffected by transformations on the original data [5–20]. An alternative approach is data augmentation [21], which expands the training set by applying the transformation to the original data multiple times, relieving the bottleneck of a small training set and allowing the network to infer the symmetry.

Both of these approaches, however, assume that the symmetry is exact. If instead, the symmetry is only approximately respected, neither approach is entirely appropriate. Imposing the exact symmetry as a constraint on the functional space will exclude the optimal point in the full space, limiting performance on the task. Augmenting the training data with examples that respect the symmetry exactly, unlike the true examples, will train the network to solve the wrong problem. In real-world scenarios, approximate symmetries are the norm and exact symmetries the exception. For example, convolutional networks assume translational invariance of images, but discrete pixelization slightly breaks this symmetry [22]. A pixelated image of a translated cat does not exactly match the translated image of a pixelated cat. For calorimeter images in particle physics, the pixel sizes are large and nonuniform, further exacerbating this issue [23]. Similarly, while theories of particle physics are Lorentz invariant, the detectors are not symmetric and are at rest in one specific frame; their measurement resolution and response thresholds are not invariant to a change of particle momentum in that frame.

In previous work, we introduced a technique to allow for data augmentation in such cases [23], which requires potentially expensive regeneration of the augmented examples before the symmetry is broken. In this paper, we outline a strategy to tackle approximate invariance by encoding the approximate symmetry directly into the network structure, which does not require resimulation. This is shown for a task for identifying collision events where a $Z$ boson decays to two muons from background processes, where a Lorentz symmetry in the data is slightly broken due to detector effects.

The organization of this paper is as follows. Section II provides the details of the dataset used, summarizing how it is generated and the structure of the resulting data. Section III covers the proposed method for encoding approximate symmetries. Section IV describes the network training and performance. Section V explores the significance of the results and summarizes the findings and future outlook.

## II. DATASET

Samples of simulated particle interactions are used to train the networks and evaluate their performance. The simulation of the detector response is simplified to allow for clear demonstration of the effect, while capturing the essential behavior of real detector systems.

The dataset describes the decay of a $Z$ boson into a pair of muons. Detecting the muons and measuring their momenta and direction allow for reconstruction of the invariant mass of the $Z$ boson despite never observing it directly. The invariant mass serves as a powerful discriminator against background processes that generate pairs of muons without an intermediate heavy particle, leading to a smooth mass spectrum.

As the true mass is Lorentz invariant by construction, the distribution of reconstructed mass should peak near the true mass regardless of the transverse momentum of the $Z$, or the muons. However, the resolution of the detector systems is not invariant to changes of the muon momenta, but depends on the transverse momentum itself, leading to a momentum dependence of the *width* of the reconstructed $Z$ mass distribution. The dependence originates from the underlying measurement technique. The magnetic field is parallel to the $z$ (beam) axis and the particle curvature is exclusively in the $x-y$ plane. The momentum of a muon is deduced from measurement of the radius of curvature of the particle's path in a magnetic field reconstructed from space-points reported by the tracking detectors. Measurement of those space points has inherent uncertainty due to their finite size, which results in an uncertainty in the measured radius which is proportional to the radius itself. Intuitively, it is harder to distinguish differences in larger radii from a small portion of the helix than in smaller radii, where a larger fraction of the helix is observed. Muon momentum resolution is therefore significantly poorer at high muon $p_{\mathrm{T}}$, which results in a poorer mass resolution for the reconstructed $Z$ boson. Therefore, while the underlying $Z$ boson mass is of course invariant, the resolution of its measurement is not. This leads to a classification task which depends on the momentum of the muon, which in a general case also depends on the momentum of the $Z$ boson, in a way that does not respect the underlying Lorentz symmetry.

Simulated collisions of protons at $\sqrt{s} = 13$ TeV resulting in $Z$ bosons and their decay to muons are generated MadGraph5 v3.5.7 [24], corresponding to the process,

$$pp \rightarrow Z \rightarrow \mu^+\mu^-.$$

Simulated collisions without an intermediate $Z$ are generated similarly to serve as the background process,

$$pp \rightarrow \gamma \rightarrow \mu^+\mu^-.$$

No initial or final state radiation is included, giving the muon pair no transverse momentum, but individual muons have sufficient spread in transverse momentum for this demonstration. The distributions of muon transverse momenta in these samples are not identical due to artifacts of the generation process. To avoid learning the details of the muon $p_{\mathrm{T}}$ distribution, background events are reweighted in bins of muon $p_{\mathrm{T}}$ so that their distribution matches the signal. The $p_{\mathrm{T}}$ distribution was split into 100 bins for this process, but is demonstrated with 500 in Fig. 1. The probability of sampling an event is determined by the assigned weight. The small differences between the signal and "reweighted signal" distributions seen are purely due to the randomness in this sampling procedure.

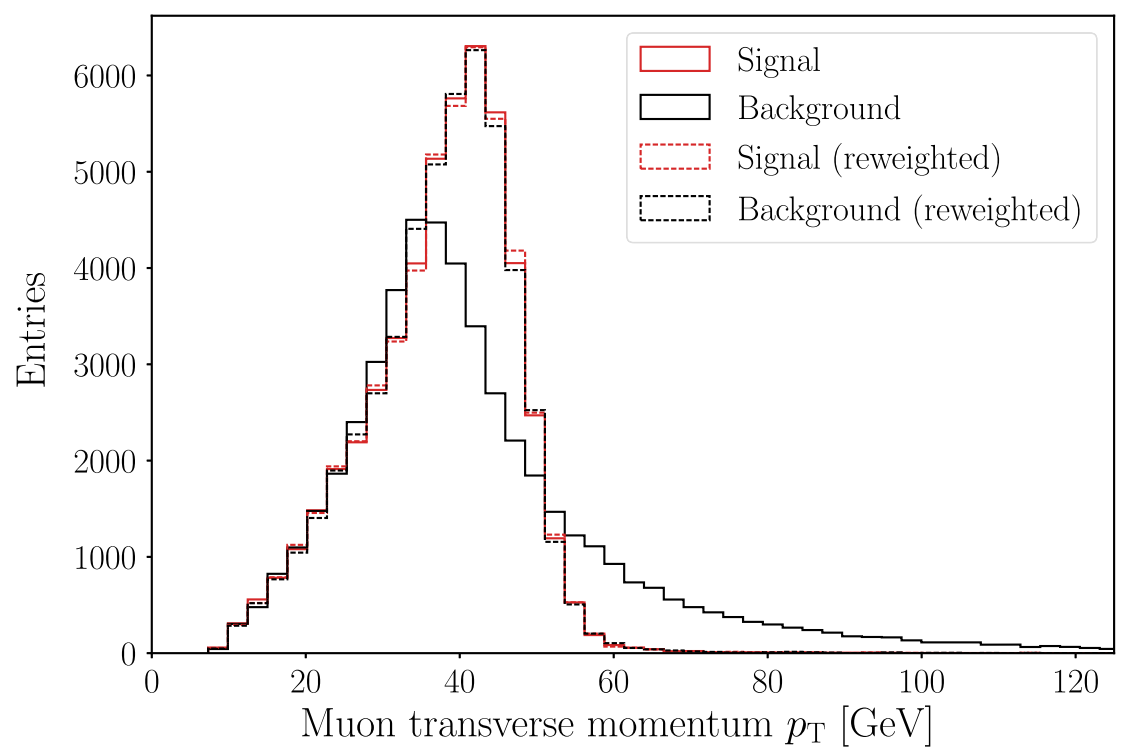

FIG. 1. Demonstration of background reweighting to match signal muon $p_{\mathrm{T}}$ distribution using 500 bins. Differences between the signal and "reweighted signal" distributions are purely due to the randomness in this sampling procedure.

When reweighting needs to be performed in higher dimensions, LHC experiments often use multivariate methods such as with neural networks [25,26], but a histogram-based reweighting was sufficient in this case.

Transverse momentum detector resolution is modeled with a parametric approach, where individual muon measured momenta are derived from the true momenta and a Gaussian smearing term, with width dependent on muon $p_{\mathrm{T}}$. We choose the width to be 10% of the true $p_{\mathrm{T}}$, such that $\sigma = p_{\mathrm{T}}^{\mathrm{true}} * 0.1$.

$$p_{\mathrm{T}}^{\mathrm{meas}} = p_{\mathrm{T}}^{\mathrm{true}} + \mathcal{N}(\mu = 0, \sigma = p_{\mathrm{T}}^{\mathrm{true}} * 0.1).$$

This is somewhat larger than expected at current LHC experiments, but helpful to exaggerate the effect for demonstration purposes. The $p_{\mathrm{T}}$-dependent effects of the reconstructed mass resolution are shown in Fig. 2. The top pane shows the original $Z$ boson mass distributions before muon $p_{\mathrm{T}}$-dependent smearing. As expected the distributions show no dependence on muon $p_{\mathrm{T}}$. The bottom pane shows the $Z$ boson mass distributions after the muon $p_{\mathrm{T}}$-dependent smearing, demonstrating the new dependence of the mass on the muon $p_{\mathrm{T}}$, which introduces a $p_{\mathrm{T}}$-dependence in the classification task. A classifier which insists on complete Lorentz invariance must apply the same strategy across all $p_{\mathrm{T}}$, which cannot be optimal.

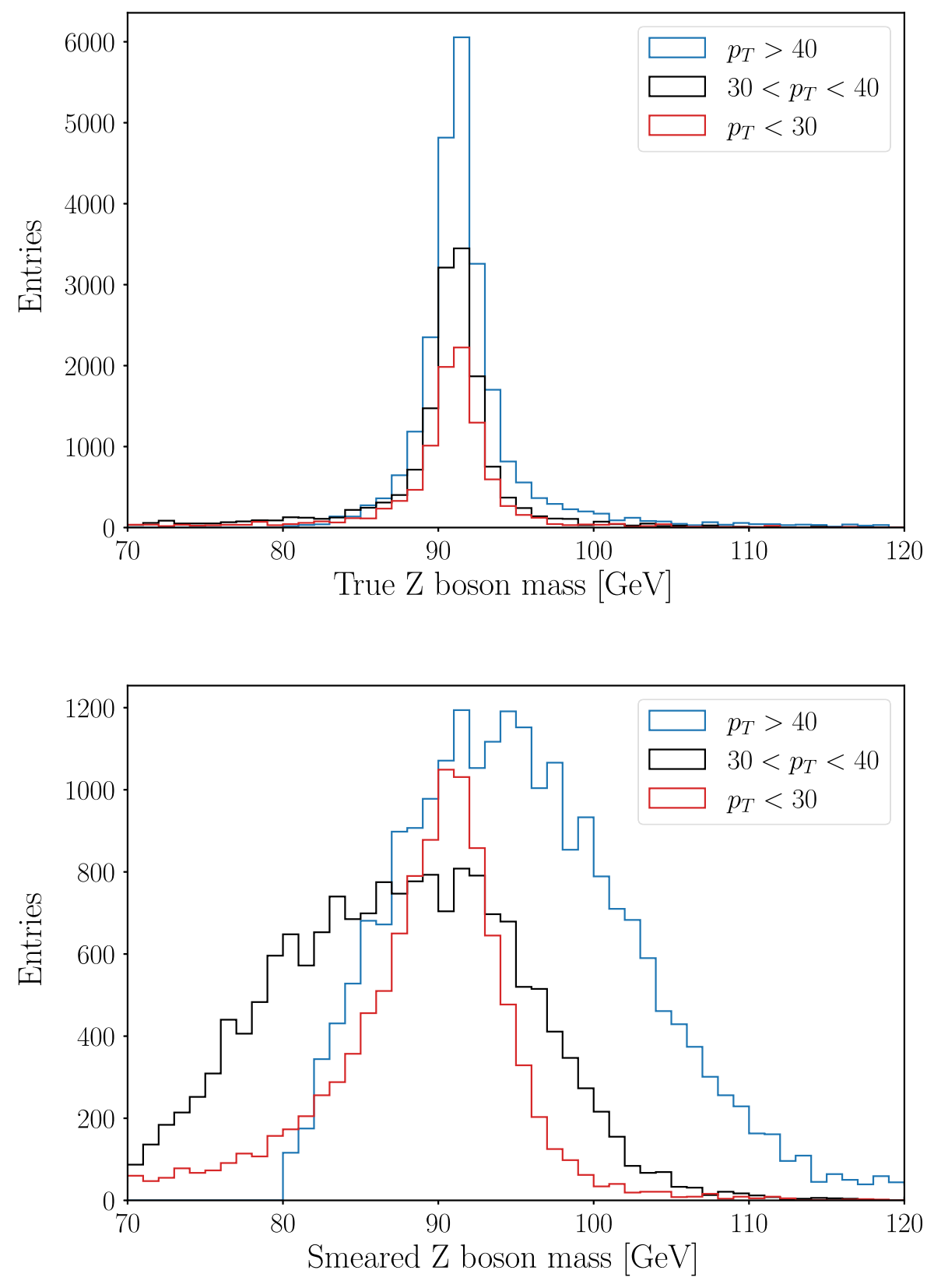

FIG. 2. Top: mass distribution for signal in three $p_{\mathrm{T}}$ slices. Bottom: same after $p_{\mathrm{T}}$-dependent smearing.

## III. APPROACH

The signal-background classification task described above is not invariant to the momentum of the $Z$, due to momentum-dependent resolution effects. Though the underlying interaction is Lorentz invariant, the data themselves are only approximately invariant, and the classification task is not invariant.

Previous work to exploit symmetries to improve network learning efficiency [5–17] has relied on techniques that allow the encoding of a perfect symmetry through the construction of a set of invariant input quantities, or by modifying the network structure to enforce invariance or equivariance under the symmetry's transformations. DeepTopLoLa [17], for example, constrains the network to work with learnable linear combinations of input four-vectors, before reducing them to a selected set of physical observables, which includes Lorentz invariants. These constraints enable such networks to learn more rapidly than more general networks. However, in the case where the symmetry is only approximately respected, they are not capable of describing the ideal solution.

We aim to capture the rapid-learning benefits of the invariant network while evading the ceiling introduced by imposing the exact symmetry. Earlier efforts [22] attempt to solve the coordinate dependence by explicitly adding the grid definition to their convolutional layers. Another approach is a soft convolutional inductive bias via gated positional self-attention [27], which focuses on a particular symmetry of a specific layer and so is somewhat less general. In reinforcement learning, residual pathway priors have been used to enforce soft priors which encourage use of inductive bias without strictly enforcing them [28]. In particle physics, some networks enforce Lorentz-invariance but break the symmetry [29,30] by specifying the lab frame or the beam axis. Most similarly, semiequivariant networks were constructed [31] to balance generality and equivariance, but the mixture was fixed for each network, rather than a learned function of the data.

Our model has two subnets, one agnostic to the symmetry, provided full muon four-vectors as inputs, and the other constrained by the exact symmetry, taking a set of invariants calculated from the same four-vectors as inputs. The input quantities supplied to the invariant network are calculated just as they would be if the symmetry were perfectly preserved, in which case they would be ideal summary statistics. However, as the symmetry is only approximate in reality, some information useful to the problem as it is represented in the data may be lost in their computation. While the four-vectors do not suffer from this limitation, their higher dimensional nature can lead to less efficient convergence during network training. By providing both representations to the network, the network can leverage the useful physical structure of the invariant quantities to better facilitate convergence, without sacrificing useful discriminating information.

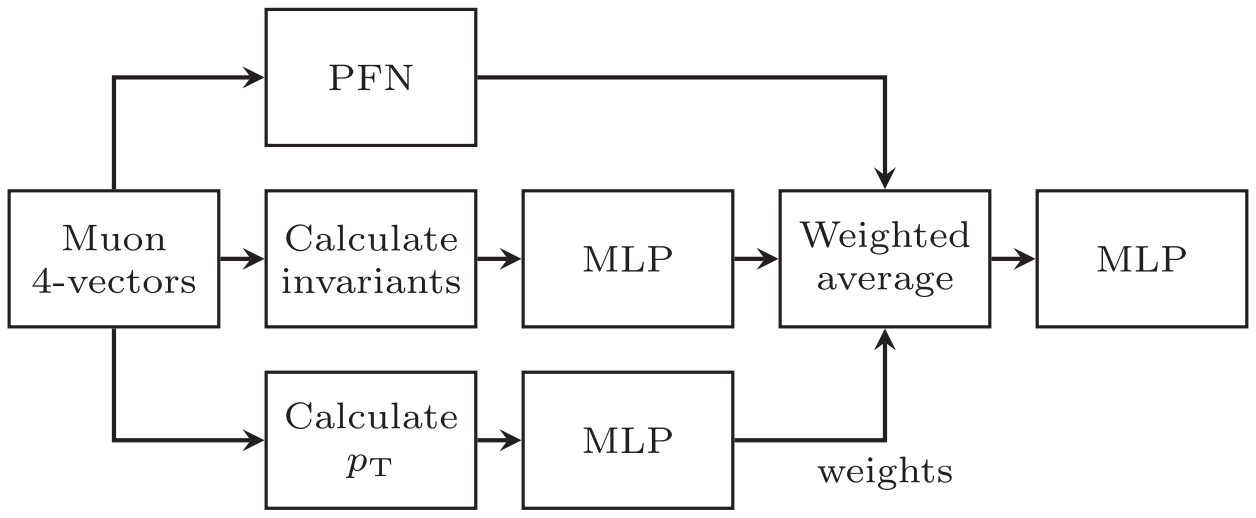

FIG. 3. Model used to learn approximate Lorentz-invariant classifier (hybrid model). The PFN subnet is unconstrained, while in parallel a Lorentz-invariant classifier learns a constrained solution. The two outputs are averaged using a weight calculated by an MLP, which simultaneously learns a $p_{\mathrm{T}}$-dependent weighting function.

The symmetry agnostic network is implemented as a particle flow network [32] (PFN), while the invariant network is implemented as a fully-connected feed-forward network (denoted here as MLP). The invariant network utilizes the squared Minkowski norm and the Minkowski inner product to preserve Lorentz symmetry. The two subnets are combined using a *learned* $p_{\mathrm{T}}$-dependent symmetry factor, which weights the general and invariant network depending on the $p_{\mathrm{T}}$ context. Figure 3 shows the structure of the model. For example, the learned-weight MLP could receive an arbitrary number of parameters. A PFN-like subnet could even be used for this task. We chose to explicitly encode the relationship between symmetry breaking and $p_{\mathrm{T}}$ due to our knowledge of the $p_{\mathrm{T}}$-dependence of the symmetry, which allows for the clearest insight into the performance for this simple example.

Each subnet produces a latent space representation of the event. Then, the two representations are combined through weighted average pooling, where the learned symmetry factor gives the weight. For a given latent space dimension $l$, learned-weight $w$, and latent space representations $\boldsymbol{g}$ and $\boldsymbol{s}$, for the general and symmetric networks, respectively, the final latent space representation $\boldsymbol{x}$ is

$$\begin{bmatrix} x_1 \\ x_2 \\ \vdots \\ x_l \end{bmatrix} = w \begin{bmatrix} g_1 \\ g_2 \\ \vdots \\ g_l \end{bmatrix} + (1-w) \begin{bmatrix} s_1 \\ s_2 \\ \vdots \\ s_l \end{bmatrix}.$$

This is then passed through a final MLP to classify the event. For example, if the hybrid model is given an event where $p_{\mathrm{T}} \approx 40$, the weight would be around 0.6, meaning that 60% of the final latent space representation comes from the general PFN subnet. This hybrid method is compared to a general network (PFN) and an invariant network with a similar number of trainable parameters. Ensembles of networks were trained to estimate the average performance and standard deviation for each method.

## IV. NETWORK TRAINING AND PERFORMANCE

Each model was trained for a total of 100 epochs. All models were trained with binary cross entropy with logits loss, AdamW optimizer [33], a weight decay parameter of 0.01, and a batch size of 128. For both the hybrid model and the PFN, the latent space representation dimension was chosen to be $l = 16$. All performance values were derived after hyperparameter tuning to maximize the area under the receiver operator characteristic curve (AUC) and minimize training volatility. During our manual search, the batch size, learning rate, weight decay, and latent space dimension size were varied. We updated these hyperparameters until the AUC increased smoothly during training and then fine-tuned them for further performance. The batch size was varied between $2^3$ and $2^7$ in powers of two. The learning rate was varied in powers of ten from $10^{-6}$ to $10^2$, with deviations by a factor of five to explore promising parts of the search space. The weight decay rate was varied between $10^0$ and $10^{-3}$ in powers of ten. The latent space dimension size was varied between $2^1$ and $2^5$ in powers of two. All models were implemented in PyTorch [34] and trained on an Intel Core i7 6-core processor (2.6 GHz).

For fair comparisons, each model employed roughly the same number of neurons across all subnets. We chose for each model to have a sum total of 1152 hidden layer neurons total. As a consequence of this constraint, each subnet of the hybrid model had fewer neurons than the fully invariant or PFN networks. Specific implementation details can be found in Appendix B.

To explore the learning behavior in small or large datasets, we restrict the training to 20% of the original full sample, or use the full sample, respectively. This is then split into training and testing sets, with 20% reserved for testing. We compare the performance over time of our hybrid approach to a general network and an invariant network; see Fig. 4. As hypothesized, the constraints of the invariant network allow it to learn more rapidly than the general classifier, but prevent it from reaching the same asymptotic performance. The hybrid offers the best of both worlds, the rapid learning typical of symmetric networks without their performance ceiling.

The balance of the two subnets is a learned behavior here, and Fig. 5 shows that as $p_{\mathrm{T}}$ grows and the symmetry is more dramatically violated, the hybrid network relies more on the general network. The brief fluctuation in the learned weight between $p_{\mathrm{T}}$ 40 and 50 is due to the high concentration of events in that range (see the histogram in Fig. 1). The general subnet can better differentiate between events as it has access to more metrics, so it is specifically necessary on this small $p_{\mathrm{T}}$ range with so many samples. Note that, in one out of the five trials, the learned weight MLP converged on a very different shaped function. In this minority case, the invariant subnet was surprisingly preferred at high $p_{\mathrm{T}}$. While the shape was unexpected, the great variation in subnet weighting for different $p_{\mathrm{T}}$ contexts shows that the hybrid model learns the broken symmetry, even when its reasoning is unclear. This fringe case performed comparably with the majority of trials.

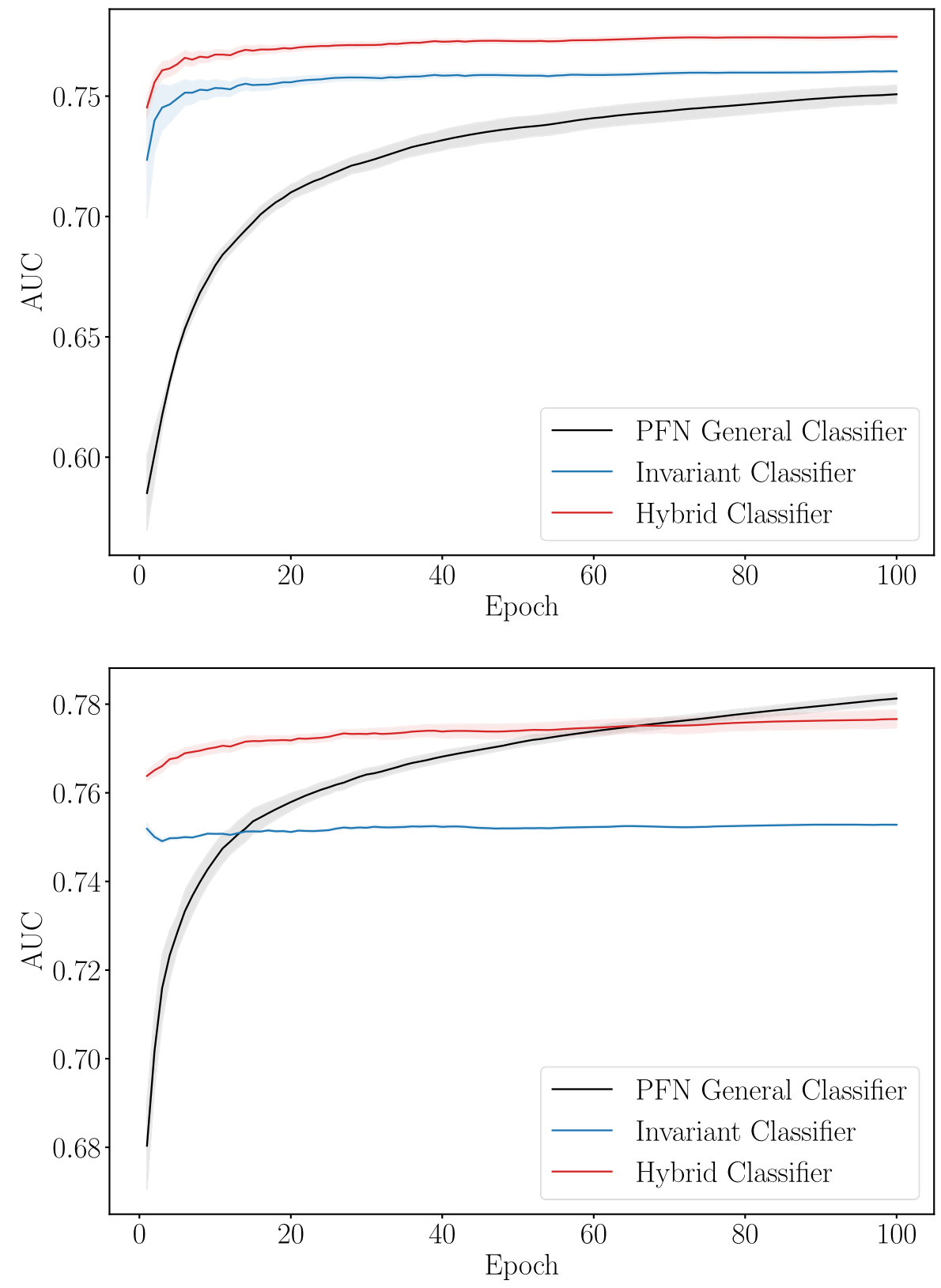

FIG. 4. Network performance, measured by AUC, versus training epoch for the three types of classifiers in scenarios with a small (top) or large (bottom) dataset. Colored bands indicate the degree of statistical variation and correspond to one standard deviation across five ensembles.

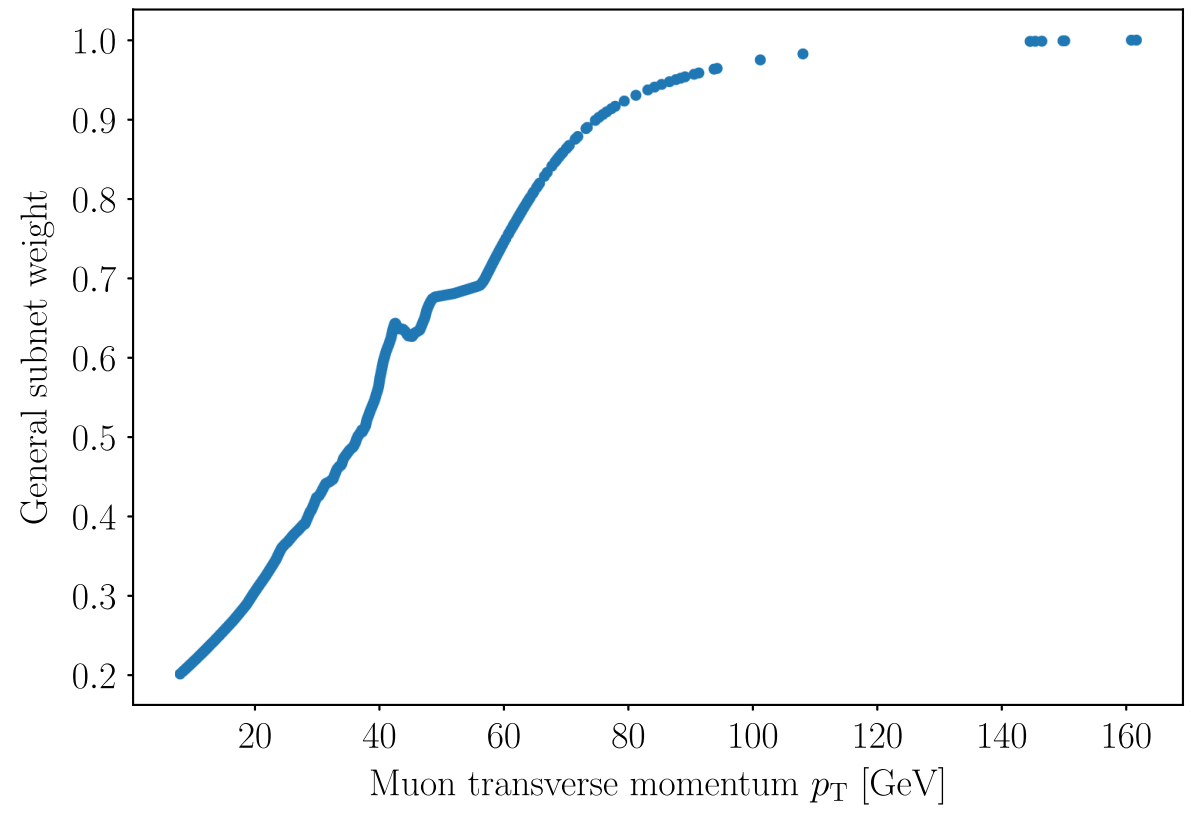

FIG. 5. Proportion of the general network used in the hybrid classifier, versus transverse momentum, showing the learned increased reliance on the general network at high momenta. Model trained on 20% of total dataset size (20,000 samples).

Note also that the general network slightly outperforms the hybrid network for the full datasets and after long training times, where the limitation is from the network capacity, and the hybrid's general subnet has one-quarter of the parameters of the fully general network. When the hybrid subnet size equals the general network size, their asymptotic performance converges (see Appendix A). Diverting some capacity to the invariant network boosts the convergence rate dramatically, but when data are plentiful, a more general classifier can always learn the patterns. A similar behavior emerges when one trains on a varying fraction of the dataset, see Fig. 6.

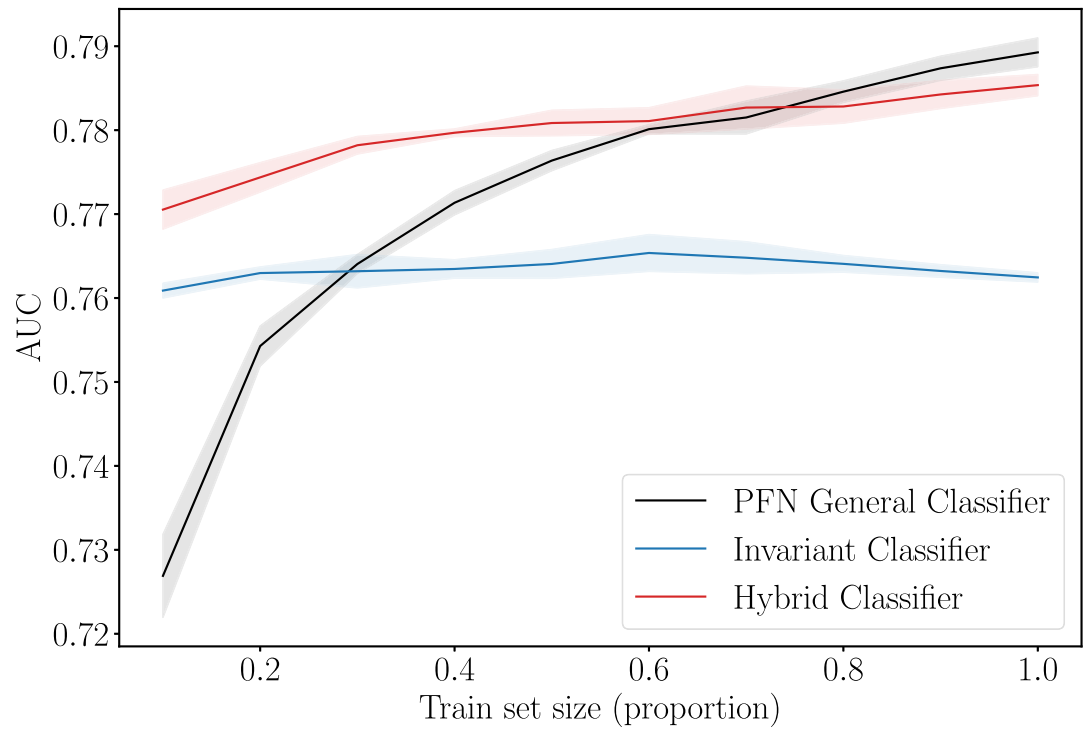

FIG. 6. Performance versus fraction of dataset used in training for the three types of networks. Colored bands indicate the degree of statistical variation and correspond to one standard deviation across five ensembles.

## V. CONCLUSIONS

Imposing constraints on networks that are respected in the training data is a powerful technique to boost rate of convergence, often allowing much greater performance with smaller training sets. However, in most realistic tasks, constraints are only approximately respected, and insisting on absolute constraints can impose a ceiling on performance. We propose a learning model which balances the generality and asymptotic performance of unconstrained networks with the rapid learning of constrained networks, using a learned symmetry factor. In a simplified toy example that demonstrates violation of Lorentz invariance, our model learns as rapidly as symmetry constrained networks but escapes its performance limitations. In problems with limited training data where symmetries are not exactly respected, our method of learning an approximate symmetry can provide significant performance boosts relative to networks that impose an exact symmetry, and can learn much more rapidly than a general, unconstrained network. In the dataset used for this study, the degree of symmetry breaking is known to depend on the transverse momentum, allowing this relationship to be explicitly incorporated into the hybrid model architecture. This architecture design enables the visualization of each subnetwork's contribution to the final decision of the hybrid network as a function of the transverse momentum (Fig. 5).

However, a hybrid network architecture—comprising both an invariant subnetwork and a general subnetwork—can also be designed without explicitly embedding the $p_T$ dependence. This can be achieved by concatenating the outputs of the two subnetworks and passing the combined result to the subsequent MLP (see Appendix C).

These developments open several promising areas for future studies, including application to image processing due to pixelization or edge effects and more complex dependence of symmetry factor on the data itself. A similar architecture to the one we present could be applied to image classification to enable data-efficient and computational-efficient training of such networks. Applications of this approach to public particle physics datasets and in collider physics experiments are also promising future directions. As more technology relies on machine learning networks to process photos, videos, and other data with broken symmetries, this hybrid approach could have widespread implications.

## ACKNOWLEDGMENTS

The authors thank Sebastian Wagner-Carena, David Miller and Jesse Thaler for useful discussions and inspiring words. D. W. and A. G. are supported by The Department of Energy Office of Science.

## DATA AVAILABILITY

The data that support the findings of this article are openly available [35].

## APPENDIX A: HYBRID SUBNETS

For a large dataset, the general PFN outperforms the hybrid model on final AUC. Our fairness restraints (see Sec. IV) mean that the hybrid model has fewer learnable parameters to learn the broken symmetry, so a fully general model with sufficient data can always do better. To confirm this as the cause, we enlarged the PFN subnet in the hybrid model to match the general model. When both the subnet and the model have two hidden layers of size 128 each, their performance with the large dataset is asymptotically equal; see Fig. 7. After 100 epochs, the difference in AUC between the general and enlarged hybrid models was $0.000 \pm 0.002$.

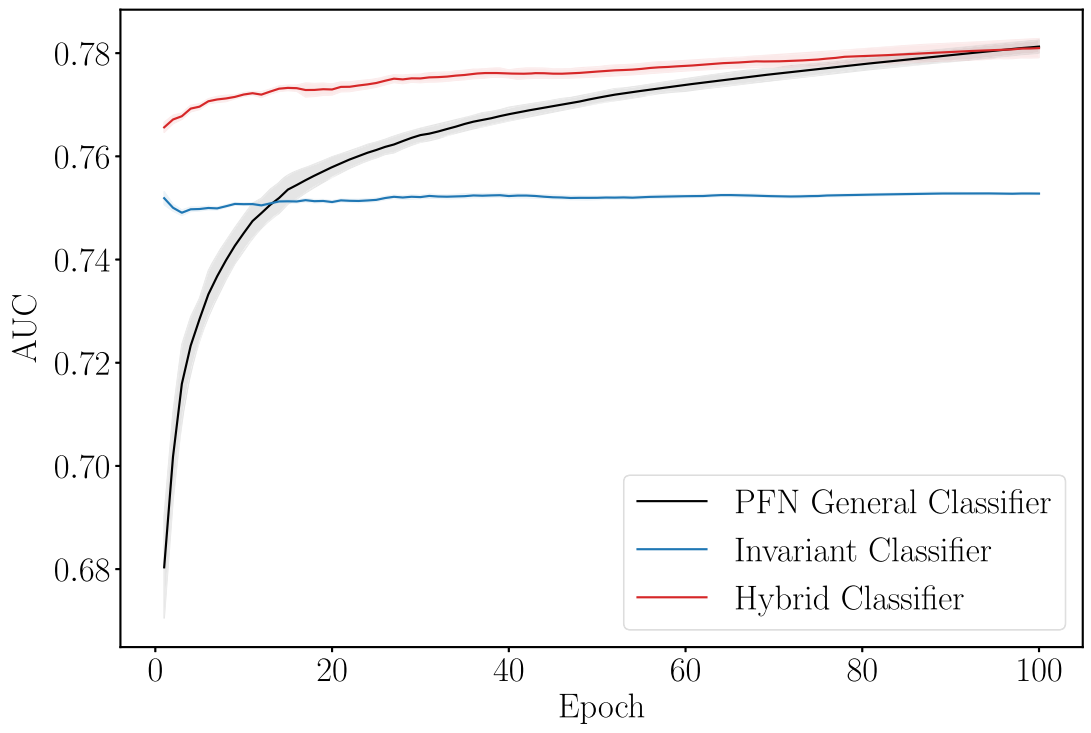

FIG. 7. Network performance on the large dataset, measured by AUC, versus training epoch for the three types of classifiers, where the PFN subnet of the hybrid model has been enlarged to match the standalone PFN size. Colored bands indicate the degree of statistical variation and correspond to one standard deviation across five ensembles.

## APPENDIX B: MODEL DETAILS

Across every model, each nonoutput Linear layer was followed by a BatchNorm1D layer and a ReLu activation function. In addition to improved performance, the batch normalization aided consistency in training and reduced fluctuations in training metrics. To preserve fairness in comparisons across networks, each model has a sum total of 1152 hidden layer neurons across all subnets. In the following model details, semicolons in between layer sizes indicate separate hidden layers.

Specific architecture details for PFNs are described in [32]. Generally, a PFN is composed of a "particle mapping" segment, followed by a classification segment implemented as an MLP. The "particle mapping" segment of the PFN has four input neurons corresponding to each of the four-momenta elements, and hidden layer sizes of 128; 128. This segment outputs a latent space representation of each final-state muon, with dimension $l = 16$. The latent space representations are summed together and passed through a classification MLP with hidden layer dimensions 512; 256; 128. The final output layer is of size 1 and has no activation function, as a sigmoid function is included in the binary cross entropy with logits loss function.

The invariant network is decidedly simpler. The four-momenta of the particles is used to compute two Lorentz-invariant quantities: the squared Minkowski norm and the Minkowski inner product. These are then passed through a single MLP with hidden layer dimensions 512; 256; 256; 128. As with the PFN, the output layer is of size 1 and has no activation function.

Finally, the hybrid network is a combination of the two baseline models. First, the general subnet is implemented as a PFN without the classification segment, only the mapping segment, using a hidden layer size of 64. Second, the invariant subnet is of similar design to the invariant network, with the difference being that it only uses a single hidden layer size of 64. Both subnets produce a latent space representation of size $l = 16$. Next, the learned symmetry factor is found by first calculating the $p_T$ values. These are then passed through an MLP with hidden layer sizes 64; 64 and an output layer of size 2, followed by a Softmax activation function. These weights are used for weighted average pooling to combine the two latent space representations produced by the subnets. Finally, this amalgam latent space representation is passed through a classifier MLP with hidden layer dimensions 512; 256; 128. The output layer is of size 1 and has no activation function.

## APPENDIX C: ALTERNATE HYBRID ARCHITECTURES

The advantage of our hybrid architecture is that the learned-weight MLP demonstrates how the network learns to balance the two subnets. By analyzing this MLP (see Fig. 5), we understand that the final latent space representation postpooling is composed more of the general subnet's output for high $p_{\mathrm{T}}$, but more of the invariant subnet's output for low $p_{\mathrm{T}}$. This shows that the network learns that the symmetry is violated more as the $p_{\mathrm{T}}$ increases. In addition, as opposed to architectures that skip weighted average pooling and use a larger input size for the classifier MLP, our architecture can be better generalized to high latent space dimensions and more subnets.

We tested multiple approaches where the latent space representations from the two subnets are pooled through concatenation instead of summation. Each pooling method was tested on the small dataset (20,000 events) for 100 epochs; the results are shown in Fig. 8. Simply concatenating the post-weighted latent space representations performed worse than the summed version. Concatenating the unweighted latent space dimensions and adding the learned weight as an additional input to the final MLP was comparable to the weighted average pooling method. The hybrid network performed best when using $p_{\mathrm{T}}$ as the additional parameter instead of the learned weight. However, this approach precludes the insights provided by the separate learned-weight function. This function enables us to understand how the hybrid network handles broken symmetries.

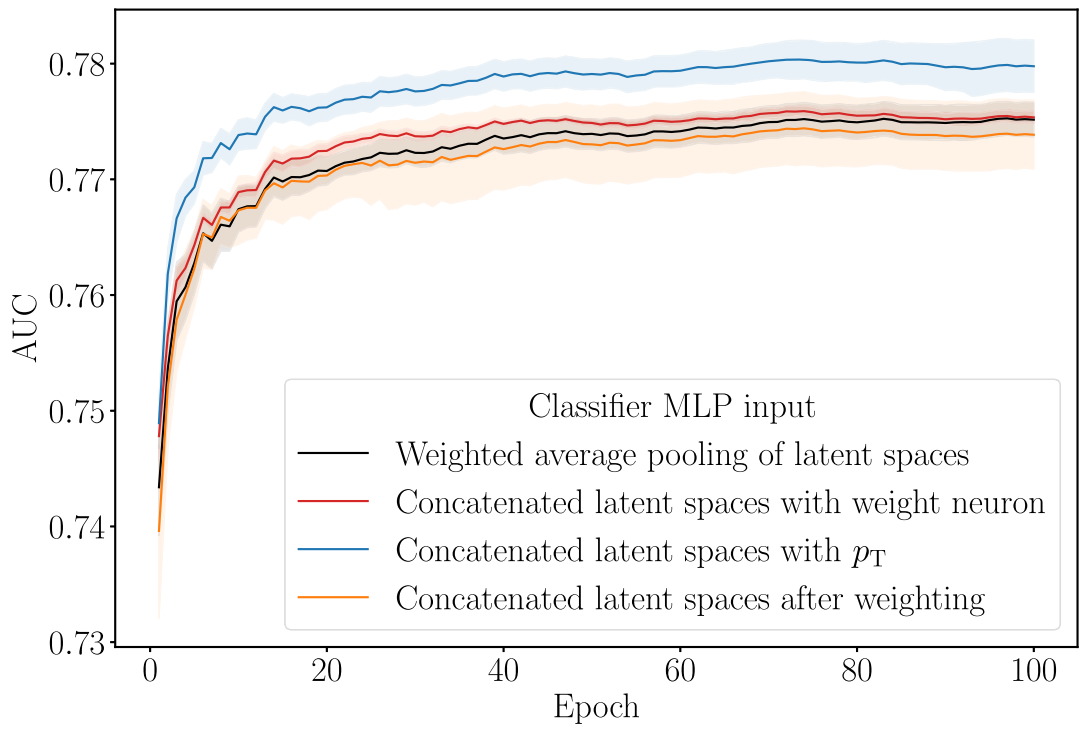

FIG. 8. Hybrid network performance on the small dataset (20,000 events), measured by AUC, versus training epoch for different pooling methods. Colored bands indicate the degree of statistical variation and correspond to one standard deviation across five ensembles.